# Realization of superchiral surface lattice resonances in three-dimensional bipartite nanoparticle arrays

Joshua T.Y. Tse and H.C. Ong[a]

Department of Physics, The Chinese University of Hong Kong, Shatin, Hong Kong, People's Republic of China

Optical chirality (OC) is a fundamental property of electromagnetic waves that plays a key role in governing chiral light-matter interaction. Here, we demonstrate how to obtain superchiral surface lattice resonances (SLRs), which arise from the hybridization between localized surface plasmons (LSPs) and diffractive Rayleigh anomalies (RAs), in nanoparticle arrays. We first study the coupling constants between LSPs and RAs in 2D Au monopartite nanorod arrays by angle-resolved reflectivity spectroscopy and finite-difference time-domain (FDTD) simulations. The complex dispersion relations and the near-fields of SLRs are then analyzed by temporal coupled-mode theory (CMT) for formulating the dependence of the coupling constants on the dipole orientation of the nanorod. The TE and TM coupling constants are found to depend strongly on the orientation of the dipole lying in the plane perpendicular to the propagation direction of RA. By using two orthogonally oriented nanorods, TE- and TM-SLRs can be excited independently. We then extend the CMT approach to rationally design superchiral SLRs based on 3D bipartite nanorods. The OC is shown to depend on the relative displacement between two nanorods, the near-field strength, and the interplay between the coupling constants and the Q-factor of the SLR resonance. We have achieved an averaged OC of 27 times stronger than that of the circularly polarized plane wave over the entire surface, featuring large area chiral surface waves that are useful for chirality-based applications.

[a] Email: hcong@phy.cuhk.edu.hk

## I. INTRODUCTION

Since the discovery of light polarization by Fresnel in the 19$^{th}$ century, elliptically polarized lights have played a major role in physical science and engineering [1]. The helicity of light has found a wide range of applications in display technology [2], telecommunication [3], microscopy [4], and spectroscopy [5]. To name a few, circularly polarized lights (CPLs) have advanced the technology of stereoscopy, enabling a cost-effective way of visualizing objects three-dimensionally [6]. Satellites use exclusively circularly polarized electromagnetic waves for communication [7]. In addition, left and right CPLs cause different absorptions in chiral molecules, giving rise to circular dichroism (CD) [8]. In fact, CD spectroscopy has become a major tool in life sciences and chemistry, particularly pharmaceutics, for differentiating right- and left-handed enantiomers [9,10]. As only left-handed amino acids are compatible with living organisms, any mismatch in the handedness can cause undesired reactions or even catastrophes [9,10]. A good example is the thalidomide tragedy that occurred in 1957 in West Germany where thousands of babies were born with missing or abnormal limbs due to the presence of right-handed chemicals in thalidomide that was released to pregnant women as painkiller [11]. Therefore, the ability to screen the chirality of chemicals at extremely low concentration is of critical importance in further developing CD spectroscopy.

As a result, much effort has been devoted to boosting the sensitivity of CD spectroscopy. Tang and Cohen have formulated the CD absorption based on optical chirality (OC) [12]. They find the absorptions taken under left- and right-CPLs can be written as:

$$A^{L,R} = \frac{\omega}{2}\left(\alpha''|\vec{E}|^2 + \chi''|\vec{B}|^2\right) \pm G''\omega \text{Im}\left(\vec{E}^* \cdot \vec{B}\right), \tag{1}$$

where $\vec{E}$ and $\vec{B}$ are the time dependent electric and magnetic fields and $\alpha''$, $\chi''$, and $G''$ are the imaginary part of the electric polarizability, the magnetic susceptibility, and the isotropic mixed electric-magnetic dipole polarizability of the chiral matter. Particularly, the second term is associated with OC, which is defined as [13]:

$$OC = -\frac{\varepsilon_0 \omega}{2} \text{Im}\left(\vec{E}^* \cdot \vec{B}\right), \tag{2}$$

where $\varepsilon_0$ is the electric permittivity. We see from Eq. (2) that OC is zero when taken under a linearly polarized excitation but reaches the highest of $OC = \pm \varepsilon_0 \omega |\vec{E}|^2 / 2$ if CPL is used. In other words, a plane wave carrying longitudinal spin angular momentum (SAM) yields OC. More

importantly, one sees from Eq. (1) that the absorption difference, $A^L - A^R$, or CD scales with OC since the first term is often identical when taken under left and right CPL excitations [8]. In addition, for a field distribution that exhibits OC larger than that of CPLs, it is coined as superchiral and is expected to improve the CD sensitivity considerably [14-19].

Evanescent waves are good candidates for realizing superchiral fields because their intensity can be made extremely high due to the strong field confinement [20,21]. However, for most of the propagating surface waves such as surface plasmon polaritons (SPPs), they carry transverse SAM instead of longitudinal and thus do not generate any OC [22,23]. Because of this, recent focus has been shifted to localized surface plasmons (LSPs) where local field helicity is found [24-28]. In fact, almost all the studies reported so far on superchiral fields employ localized resonances because they support various multipoles such as electric and magnetic dipoles and quadrupoles for assembling complex nonorthogonal electric and magnetic fields [29]. Unfortunately, the helicity sometimes is so localized at the not-so-good regions that it may not interact with molecules at all despite its presence [30,31]. It is essential to realize delocalized superchiral surface waves that can cover as much surface as possible for maximizing the chiral sensing efficiency. Effort on researching propagating surface waves that carry longitudinal SAM or strong OC is sparse compared to the LSP counterparts. Tang and Cohen use two counterpropagating CPL with opposite handedness to create a standing wave that produces superchiral field [12]. They subsequently demonstrate the effectiveness of using strong OC in discriminating the enantiomers [14]. On the other hand, Pellegrini et al excite both TM- and TE-polarized Bloch surface waves on 1D photonic crystals [32]. As the orthogonal waves have the same frequency and momentum, they overlap to form chiral surface waves. Recently, Goerlitzer et al [33] and Mattioli et al [34] have worked on chiral surface lattice resonances (SLRs) in near and mid infrared spectral regions based on 3D crescent and 2D slit arrays. SLRs are collective excitation of periodic arrays of nanoparticles by couplings the LSPs to diffractive TE- and TM-Rayleigh anomalies (RAs) [35-38]. However, all these methods mentioned so far have stringent requirements in design and implementation. In addition, there is a lack of systematic guidelines in rationally designing photonic systems for maximizing OC.

Here, we attempt to realize superchiral SLRs by using 3D bipartite nanoparticle arrays. Our approach is divided into two parts. First, we study the coupling constants between the LSP and RAs from 2D Au monopartite nanorod arrays by using angle-resolved reflectivity spectroscopy, finite-difference time-domain (FDTD) simulations, and temporal coupled mode theory (CMT) to understand the dependence of the TE and TM coupling constants on the dipole moment induced by the nanorod. Our results show the coupling constants depend on the orientation of the nanorod that lies in the plane normal to the propagation direction of RA. While TE-SLR is excited solely by a horizontal nanorod, a vertical nanorod dominates the excitation of TM-SLR. Therefore, both the magnitudes and the phases of the TE- and TM-SLRs can be controlled independently by using two orthogonally oriented nanorods. Second, we demonstrate the SLRs supported by a bipartite system are superchiral. The relative displacement between two nanorods, the near-field strength, and the interplay between the coupling constants and the Q-factor of the SLR are shown to play a key role in determining OC. By properly engineering these parameters, we have achieved the strongest and average OC to be 102 and 27 times stronger than that of the CPL.

## II. EXPERIMENTAL AND NUMERICAL RESULTS OF MONOPARTITE ARRAYS

We have prepared 2D square lattice Au monopartite nanorod arrays on glass substrate by using electron-beam lithography and some of the scanning electron microscopy plane-view images are shown in the insets of Fig. 1, indicating the lattices all have period $P = 400$ nm and the cuboid-like nanorods have length $L$, width $W$, and height $H = 100$, 50, and 50 nm. The nanorods are azimuthally rotated, with respect to the $\Gamma$-Y direction, from $\rho = 0°$ to $90°$ with a step size = $15°$. After sample preparation, they are then transferred to a Fourier-space optical microscope for polarization- and angle-resolved reflectivity spectroscopy [40]. Since the samples are immersed in refractive index matching oil that matches with the glass substrate, the arrays are present in a homogeneous environment.

We conduct 45° linearly polarized specular reflectivity spectroscopy on the samples taken along the $\Gamma$-X direction at different incident polar angles $\theta$. The contour reflectivity mappings, also known as dispersion relations, of $\rho = 0°$, $30°$, $60°$, and $90°$ arrays are presented in Fig. 1 [39], showing the presence of longitudinal LSP at $\lambda \approx 875$ nm and the dispersive (–1,0) RAs as the dashed lines. The dashed lines are calculated by the phase matching equation $(n/\lambda)^2 = (n\sin\theta/\lambda - 1/P)^2$, where $n = 1.5$ is the refractive index of the environment [40]. We observe an avoided crossing at $\theta \approx 23.6° - 30°$ due to the coupling between the LSP and the RAs, forming an energy band gap as well as two upper and lower coupled bands, as shown by the solid lines [41,42]. By convention, the coupled bands that follow closely with the RAs are labelled as the dispersive (–1,0) SLRs because they exhibit narrow linewidth or high Q-factor. Since the avoided crossing is not present at $\rho = 90°$, the LSP and RA are not coupled to form the coupled bands. We then extract the spectral positions $\lambda_{1,2}$ and the linewidths $\Delta\lambda_{1,2}$ of two bands from each sample by using Fano function fitting and plot them in Fig. 2(a) – (d) [38]. The energy band gaps now become apparent from the $\lambda_{1,2}$ plots. At the same time, the linewidths cross exactly where the band gap is formed. We see the size of the gap and the linewidth at the cross-point decrease with increasing $\rho$. The gap eventually disappears when $\rho = 90°$ and LSP and RA instead of the coupled bands are observed. As the gap size indicates the coupling strength, the coupling between LSP and RA weakens when $\rho$ increases [42]. To verify our results, we have performed FDTD simulations on the 2D Au rectangular cuboid arrays taken under the same conditions and the $\lambda_{1,2}$ and $\Delta\lambda_{1,2}$ plots are shown in Supplementary Information [39]. The simulation agrees well with the experiment, indicating there exists a strong dependence of the coupling strength on the orientation of the nanorod.

### III. TEMPORAL COUPLED MODE THEORY

To elucidate such dependence, we formulate the interactions between the LSP and two TE- and TM-RAs by temporal CMT [43-45]. The dynamics of three modes can be formulated as [40]:

$$\frac{d}{dt}\begin{bmatrix} a_{LSP} \\ a_{TE} \\ a_{TM} \end{bmatrix} = i \begin{bmatrix} \tilde{\omega}_{LSP} & \Omega_{TE} & \Omega_{TM} \\ \Omega_{TE} & \tilde{\omega}_{TE} & 0 \\ \Omega_{TM} & 0 & \tilde{\omega}_{TM} \end{bmatrix} \begin{bmatrix} a_{LSP} \\ a_{TE} \\ a_{TM} \end{bmatrix}, \qquad (3)$$

where $a_{LSP,TE,TM}$ and $\tilde{\omega}_{LSP,TE,TM}$ are the mode amplitudes and the complex frequencies of LSP, TE- and TM-RAs. The coupling constants between LSP and TE-RA and between LSP and TM-RA are defined as $\Omega_{TE}$ and $\Omega_{TM}$, respectively. Since TE- and TM-RAs are orthogonal, they are degenerate. In addition, they can be considered as nonradiative such that $\tilde{\omega}_{TE} = \tilde{\omega}_{TM} = \omega_{RA}$, where $\omega_{RA}$ is the RA frequency. As a result, the interaction between the modes is mostly near-field in nature and $\Omega_{TE}$ and $\Omega_{TM}$ are considered as real values [46]. By diagonalizing Eq. (3), we solve for the complex eigenfrequencies, $\tilde{\omega}_{1-3}$, to be:

$$\tilde{\omega}_{1,2} = \omega_{1,2} + i\frac{\Gamma_{1,2}}{2} = \frac{(\tilde{\omega}_{LSP} + \omega_{RA})}{2} \pm \sqrt{\left(\frac{\tilde{\omega}_{LSP} - \omega_{RA}}{2}\right)^2 + \Omega_{TE}^2 + \Omega_{TM}^2} \qquad \text{and}$$

$$\tilde{\omega}_3 = \omega_{RA}, \tag{4}$$

and the mode amplitudes, $a_{1-3}$, are:

$$a_{1,2} = \beta_{1,2}\left[\left(\omega_{RA} - \tilde{\omega}_{1,2}\right)a_{LSP} - \Omega_{TE}a_{TE} - \Omega_{TM}a_{TM}\right] \text{ and } a_3 = \frac{\Omega_{TM}a_{TE} - \Omega_{TE}a_{TM}}{\sqrt{\Omega_{TE}^2 + \Omega_{TM}^2}}, \tag{5}$$

where $\beta_{1,2} = \frac{1}{\sqrt{\left(\omega_{RA} - \tilde{\omega}_{1,2}\right)^2 + \Omega_{TE}^2 + \Omega_{TM}^2}}$ is the normalization constant. We see from Eq. (4) that while $\tilde{\omega}_3$ remains as RA-like, $\tilde{\omega}_{1,2}$ are the upper and lower bands. High Q-factor can be observed, for example, from the lower band when $\omega_{LSP} - \omega_{RA}$ is large [40]. Eq. (5) indicates how $a_{LSP,TE,TM}$ constitute the coupled modes, $a_{1-3}$. The RA-like $a_3$ composes of TE- and TM-RAs with the amplitudes governed by $\Omega_{TM}$ and $\Omega_{TE}$, and has no contribution from $a_{LSP}$. On the other hand, $a_{1,2}$ consist of all $a_{LSP,TE,TM}$ with the contribution from $a_{LSP}$ varying with $\omega_{RA} - \tilde{\omega}_{1,2}$ along the upper and lower bands. The corresponding electric near-fields are then given as $\left|\vec{E}_{1-3}\right|^2 = \frac{2|a_{1-3}|^2}{n^2\varepsilon_o V_{1-3}^{eff}}$, where $V_{1-3}^{eff}$ are the effective mode volumes [47].

We then construct the electric fields of the SLR based on the eigenvectors. We note $a_{LSP}$ is a spatially confined LSP mode that assumes a functional form of $C\frac{(\vec{p}\cdot\hat{r})\hat{r} - \vec{p}}{r^3}$, where $\vec{p}$ is the dipole moment, $\hat{r}$ is the position unit vector, and $C$ is a constant [48]. Assume the dipole lies in the surface plane as in our case, $\vec{p} = |\vec{p}|(\sin\rho\hat{x} + \cos\rho\hat{y})$ and $r\hat{r} = x\hat{x} + y\hat{y} + z\hat{z}$, where $\hat{x} = \hat{k}_{RA}$ and $\hat{z}$ are the unit vectors of the propagation direction of the RA and the surface normal, $\hat{y} = \hat{z} \times \hat{k}_{RA}$ and $\rho$ is the dipole orientation defined with respect to $\hat{y}$. On the other hand, both $a_{TE,TM}$ have Bloch forms, which can be expressed as $-u_{\vec{k}_{RA}}^{TE}e^{i\vec{k}_{RA}\cdot\vec{r}}\hat{y}$ and $-u_{\vec{k}_{RA}}^{TM}e^{i\vec{k}_{RA}\cdot\vec{r}}\left(i\sinh\tau\hat{x} + \cosh\tau\hat{z}\right)$ for the TE- and TM-RAs, respectively, where $u_{\vec{k}_{RA}}^{TE}$ and $u_{\vec{k}_{RA}}^{TM}$ are the periodic functions, $\vec{k}_{RA}$ is defined as $\vec{k}_{RA} = \vec{k}_{\parallel} + \vec{G}$ in which $\vec{k}_{\parallel}$ is the in-plane component of wavevector of the incident light, $\vec{G}$ is the reciprocal lattice vector and $\tau$ characterizes the longitudinal field component as $\tanh \tau = \sqrt{1 - \left(\dfrac{\lambda_{RA}}{\lambda}\right)^2}$, where $\lambda_{RA}$ is the RA wavelength. Note the TM-RA carries transverse spin SAM due to the $\pi/2$ phase difference between the $\hat{k}_{RA}$ and $\hat{z}$ components, resulting in an elliptical polarization spinning along the propagation direction [49]. As a result, the electric field of the lower band can be approximated as:

$$\vec{E}_2 = \beta_2 \sqrt{\dfrac{2}{n^2 \varepsilon_0 V_2^{eff}}} \begin{pmatrix} \left[\Pi\left(xy\cos\rho - (y^2+z^2)\sin\rho\right) + i\Omega_{TM}\sinh\tau u_{\vec{k}_{RA}}^{TM} e^{i\vec{k}_{RA}\cdot\vec{r}}\right]\hat{x} \\ +\left[\Pi\left(xy\sin\rho - (x^2+z^2)\cos\rho\right) + \Omega_{TE} u_{\vec{k}_{RA}}^{TE} e^{i\vec{k}_{RA}\cdot\vec{r}}\right]\hat{y} \\ +\left[\Pi\left(xz\sin\rho + yz\cos\rho\right) + \Omega_{TM}\cosh\tau u_{\vec{k}_{RA}}^{TM} e^{i\vec{k}_{RA}\cdot\vec{r}}\right]\hat{z} \end{pmatrix}, \quad (6)$$

where $\Pi = \dfrac{C(\omega_{RA} - \tilde{\omega}_2)|\vec{p}|}{r^5}$. We see from Eq. (6) that each component consists of the spatially localized LSP that supports a range of propagation constants and the Bloch waves that have well-defined propagation constant at $\vec{k}_{RA}$.

### IV. DETERMINATION OF $\Omega_{TE}$ and $\Omega_{TM}$

We attempt to determine $\Omega_{TE}$ and $\Omega_{TM}$ by using Eq. (4) & (6). In fact, Eq. (4) can be used to fit the experimental and simulation results given in Fig. 2(a) – (c) and the Supplementary Information for determining $\sqrt{\Omega_{TE}^2 + \Omega_{TM}^2}$. The best fits are shown as the dashed lines and the corresponding $\sqrt{\Omega_{TE}^2 + \Omega_{TM}^2}$ are plotted in Fig. 2(e) as a function of $\rho$. One sees both the coupling constants from the experiment and simulation agree well with each other and they decrease with increasing $\rho$, reaching zero when $\rho = 90°$. The result here is also consistent with the fact that the radiations from the LSPs should align with the propagation direction defined by the (–1,0) RA to collectively couple all the LSPs together [42]. Therefore, for longitudinal LSP, the nanorods should be oriented at $\rho = 0°$ to facilitate the strongest coupling between the LSP and RA. In contrast, when $\rho = 90°$ in Fig. 2(d), the radiation fields are now orthogonal to the RA propagation, resulting in zero coupling.

We then combine Eq. (6) and the FDTD simulated field patterns to deconvolute $\Omega_{TE}$ and $\Omega_{TM}$. We begin with the dipole set at $\rho = 0°$, and $\vec{E}_2$ is given as

$$\frac{\beta}{n}\sqrt{\frac{2}{\varepsilon_o V_2^{eff}}} \left( \begin{array}{l} \left[ \Pi xy + i\Omega_{TM} \sinh\tau u_{\bar{k}_{RA}}^{TM} e^{i\bar{k}_{RA}\cdot\bar{r}} \right]\hat{x} + \left[ -\Pi(x^2+z^2) + \Omega_{TE} u_{\bar{k}_{RA}}^{TE} e^{i\bar{k}_{RA}\cdot\bar{r}} \right]\hat{y} \\ + \left[ \Pi yz + \Omega_{TM} \cosh\tau u_{\bar{k}_{RA}}^{TM} e^{i\bar{k}_{RA}\cdot\bar{r}} \right]\hat{z} \end{array} \right).$$ The nanorod array taken at $\theta = 24°$ and $\lambda = 950$ nm is then simulated by FDTD and the $(-1,0)$ SLR near-fields propagating along the $\Gamma$-X direction are displayed in Fig. 3(a) – (c) for $\rho = 0°$. The field patterns show the SLR is TE-like, suggesting $\Omega_{TM} = 0$. Likewise, for $\rho = 30°$ and $60°$ at the same wavelength, the field patterns illustrated in the Supplementary Information all indicate the SLRs are TE-like or $\Omega_{TM} = 0$ [39]. On the other hand, for the $\rho = 90°$ nanorod, the field patterns in Fig. 3(d) – (f) show only localized fields are present around the nanorod and they agree with Eq. (6) that $\vec{E}_2 = \frac{\Pi\beta}{n}\sqrt{\frac{2}{\varepsilon_o V_2^{eff}}}\left[(y^2+z^2)\hat{x} + xy\hat{y} + xz\hat{z}\right]$, which is a spatially localized field if both $\Omega_{TE}$ and $\Omega_{TM} = 0$. In fact, it verifies our reflectivity measurement that $\sqrt{\Omega_{TE}^2 + \Omega_{TM}^2} = 0$ in Fig. 1(d). Therefore, it is reasonable to conclude, for the nanorod lying on the surface, $\Omega_{TM} = 0$ always regardless of $\rho$ and Fig. 2(e) indicates the variation of $\Omega_{TE}$ only.

## V. DEPENDENCE OF $\Omega_{TE}$ and $\Omega_{TM}$ ON DIPOLE MOMENT

It is then natural to raise a question whether it is possible to have nonzero $\Omega_{TM}$. To answer it, we proceed in the following to show nonzero $\Omega_{TM}$ exists only when the dipole is perpendicular to the RA propagation direction and at the same time aligned with the incident TM polarization. We notice the Bloch components in Eq. (6) can be written as:

$$\frac{\Omega_{TM}\left|u_{\bar{k}_{RA}}^{TM}\right|}{\Omega_{TE}\left|u_{\bar{k}_{RA}}^{TE}\right|} = e^{i(\phi_z-\phi_y)}\sqrt{\frac{\left|\vec{E}_2\cdot\hat{z}\right|^2_{Bloch} - \left|\vec{E}_2\cdot\hat{x}\right|^2_{Bloch}}{\left|\vec{E}_2\cdot\hat{y}\right|^2_{Bloch}}}. \quad (7)$$

where $\phi_{y,z}$ are the phases of the $y$- and $z$-Bloch components. When the nanorod is much smaller than the wavelength and both TE- and TM-RAs are excited equally, we approximate $\left|u_{\bar{k}_{RA}}^{TM}\right| \approx \left|u_{\bar{k}_{RA}}^{TE}\right|$ and Eq. (7) yields $\Omega_{TM}/\Omega_{TE}$. If one can deconvolute the Bloch components of the SLR fields in different directions, the coupling constant ratio can be determined accordingly. Consider the fact that the Bloch and localized waves in Eq. (6) show distinct momentum dependences, Fourier transform of the near-fields from real to momentum space should be useful for such deconvolution.

We demonstrate it by carrying out FDTD simulations on a unit cell as shown in the inset of Fig. 4(a) where the cuboid nanorod is rotated in the plane orthogonal to the Γ-X direction. The orientation of the nanorod is defined as $\sigma$ with respect to the surface normal such that $\vec{p} = |\vec{p}|(\sin\sigma \hat{y} + \cos\sigma \hat{z})$. A 45° linearly polarized light is illuminated on the system at $\theta = 24°$ along the Γ-X direction to ensure both TE- and TM-RAs are excited equally. The absorption spectra of the arrays for $\sigma = 0°$, 30°, 60°, and 90° are shown in Fig. 4(a) indicating (–1,0) SLRs are present at $\lambda = 868$ nm. We then calculate the corresponding SLR near-field patterns at 868 nm in Fig. 4(b) – (e). To deconvolute the Bloch components, we perform Fourier transform on the real space electric fields in different directions to obtain the magnitude and relative phase mappings in momentum space in Fig. 5(a) – (e) and (f) – (j) for $\sigma = 0°$ and 90° arrays, respectively. Results for $\sigma = 30°$ and 60° are provided in the Supplementary Information [39]. Apparently, the magnitude mappings consist of several discrete peaks superimpose on a broad background. While the localized fields contribute to the background, the peaks are Bloch waves that have well-defined Bragg scattering vectors, following well the phase matching equation. The plots of the Bloch $|E_x|^2$, $|E_y|^2$, and $|E_z|^2$ in different directions and their relative phase differences for the fundamental order are shown in Fig. 6(a) and (b) as a function of $\sigma$. We see from the plots that the SLR is no longer TE- or TM-polarized but a mix of them. In addition, although the phase difference between $E_x$ and $E_z$ is always 90°, which agrees with the functional forms of the TM-Bloch waves in Eq. (7), $E_y$ and $E_z$ are either 0° or 180° phase shifted.

Following Eq. (7), we use the data to determine $|\Omega_{TM}/\Omega_{TE}|$ and plot it in Fig. 6(c). The ratio increases by almost eight orders of magnitude when the nanorod rotates from horizontal to vertical orientation, revealing the coupling between LSP and TM-RA is the strongest when the nanorod is vertical. In addition, the ratio reaches 1 when $\sigma = 45°$ where the couplings between the LSP and TM- and TE-RAs become equal. Such $\sigma$ dependence leads us to argue the ratio is solely determined by the dipole moment projected in the z- and y-directions. In fact, knowing the dipole moment ratio is $\hat{p} \cdot \hat{z}/\hat{p} \cdot \hat{y} = \cot\sigma$, we fit $|\Omega_{TM}/\Omega_{TE}|$ in Fig. 6(c) by using $|\beta \cot\sigma|$, where $\beta$ is a constant. Remarkably, we find the ratio follows the $\cot\sigma$ dependence so well that $\beta$ is found to be 0.981. Therefore, we conclude $\Omega_{TE} \propto \hat{p} \cdot \hat{y}$ and $\Omega_{TM} \propto \hat{p} \cdot \hat{z}$, and it reinforces our earlier study that any dipole oriented along the *x*-direction, i.e. the propagation direction of the RA, plays no role in controlling $\Omega_{TE}$ and $\Omega_{TM}$. More importantly, by using dimerized horizontal and vertical nanorods, TE- and TM-SLRs can be excited independently. They can then be overlapped with proper phase shift to yield superchiral SLR [32].

## VI. DESIGN AND REALIZATION OF SUPERCHIRAL SLR IN 3D BIPARTITE ARRAYS

We are now at the position to rationally design the OC from bipartite arrays. In fact, the OC can be analytically formulated by making use of Eq. (2) & (6). We assume the interactions between two nanorods is weak due to their orthogonality and Eq. (3) is sufficient for describing each nanorod. The total electric field generated by the dimer thus is the summation of TE- and TM-SLRs expressed as $\vec{E}_{TE} + \vec{E}_{TM} e^{i\Phi}$, where $\vec{E}_{TE}$ and $\vec{E}_{TM}$ are given by $\sigma = 0°$ and $90°$ or $\Omega_{TE}$ and $\Omega_{TM} = 0$ for the horizontal and vertical nanorods, and $\Phi$ is the phase difference between two fields. If $\omega_{SP} - \omega_{TE}$ is large so that $\Pi \approx 0$, the total field becomes

$$\sqrt{\frac{2}{n^2 \varepsilon_o}} \left( \frac{\beta^{TE}}{\sqrt{V_2^{eff,TE}}} \Omega_{TE} u^{TE}_{\bar{k}_{RA}} e^{i\bar{k}_{RA}\cdot\vec{r}} \hat{y} + \frac{\beta^{TM}}{\sqrt{V_2^{eff,TM}}} \Omega_{TM} u^{TM}_{\bar{k}_{RA}} e^{i\bar{k}_{RA}\cdot\vec{r}} \left( i\sinh\tau\hat{x} + \cosh\tau\hat{z} \right) e^{i\Phi} \right). \quad (8)$$

For a special case where 45° linear polarization is used, the averaged OC can be approximated as [39]:

$$OC_{avg} = \frac{\omega}{nc} \frac{\left\langle |u_{\bar{k}_{RA}}|^2 \right\rangle}{V_{SLR}} \frac{\Omega^2}{(\omega_{RA} - \tilde{\omega}_2)^2 + \Omega^2} \sin\Phi, \quad (9)$$

where $\Omega_{TE} = \Omega_{TM} = \Omega$, $\beta^{TE}\Omega_{TE} = \beta^{TM}\Omega_{TM} = \dfrac{\Omega}{\sqrt{(\omega_{RA} - \tilde{\omega}_2)^2 + \Omega^2}}$, $V_2^{eff,TE} \approx V_2^{eff,TM} = V_{SLR}$, and $\left\langle |u_{\bar{k}_{RA}}|^2 \right\rangle$ is the average integral of the energy of $u_{\bar{k}_{RA}}$ over one unit cell provided $|u^{TM}_{\bar{k}_{RA}}| \approx |u^{TE}_{\bar{k}_{RA}}| = |u_{\bar{k}_{RA}}|$. We expect both $\Omega$ and $\left\langle |u_{\bar{k}_{RA}}|^2 \right\rangle / V_{SLR}$ depend on the geometry of the nanorods and $\Phi$ is a function of the relative displacement between two nanorods. In addition, $\dfrac{\Omega^2}{(\omega_{RA} - \tilde{\omega}_2)^2 + \Omega^2}$ displays a Lorentzian profile where the interplay between $\Omega$ and $\omega_{RA} - \tilde{\omega}_2$ is important. Finally, as the Q-factor of the SLR scales with $1/(\omega_{RA} - \tilde{\omega}_2)^2$ [40], strong OC should come along with high Q-factor.

We first verify the $\sin\Phi$ dependence. We simulate two series of $P = 400$ nm arrays with cylindrical nanorods that have $L$ and radius $R = 100$ and 25 nm. As shown in the inset of Fig. 7(b), while the center of the horizontal nanorod is always fixed at $(x, y) = (150, 0)$ nm, the bottom of the vertical nanorod is varied from $(-200, 0)$ to $(100, 0)$ and from $(-150, -100)$ to $(-150, 100)$ for the first and second series. An incident light is set at $\theta = 24°$ along the Γ-X direction to excite the (−1,0) SLRs at 877 nm. The field patterns in different directions are calculated and then Fourier transformed to momentum space for determining the phase difference, or Φ, between $E_y^{TE}$ and $E_z^{TM}$, which are the field components of the TE- and TM-SLRs. The results are summarized in Fig. 7, showing Φ decreases linearly from 135° to –135° when the $x$-direction relative displacement, $d_x$, increases but remains almost constant with the $y$-relative displacement, $d_y$. As an illustration, we simulate the OC mappings taken at 25 nm above the $xy$-plane for the two series and then calculate the averaged OC ratio over the entire unit cell in Fig. 7 after normalizing with the OC of CPL. From two figures, we see $OC_{avg} > 1$, indicating orthogonal nanorods are effective in making superchiral SLRs. In Fig. 7(a), $OC_{avg}$ varies sinusoidally from –4 to 4, in consistent with the $\sin\Phi$ dependence provided $\Phi = -2\pi d_x/P$. Likewise, the $d_y$ dependence shows $OC_{avg}$ does not change as Φ is almost constant. Therefore, the dependences reveal $d_x$ is exclusive to engineer Φ and thus the helicity of SLRs in a wide range between left and right.

Next, we study $\langle |u_{\tilde{k}_{RA}}|^2 \rangle / V_{SLR}$ dependence and the interplay between Ω and $\omega_{RA} - \tilde{\omega}_2$. We simulate the $OC_{avg}$ ratio spectra of (−1,0) SLRs from four bipartite nanorod arrays with the relative displacements $d_x$ and $d_y$ are set at –100 and 0 nm and the length $L$ is varied from 80 to 140 nm in Fig. 8(a). The corresponding absorption spectra are shown in Fig. 8(b). We see the strongest $OC_{avg}$ ratio = 27.5 is observed at $L = 80$ nm with the maximum local OC ratio = 102.5, as shown in the Supplementary Information [39]. Apparently, by comparing Fig. 8(a) & (b), one sees SLR that has narrower linewidth at the same time also exhibits stronger $OC_{avg}$. We then plot the averaged near-field intensity over the unit cell taken at the $OC_{avg}$ maximum and the Q-factor as a function of the $OC_{avg}$ maximum in Fig. 8(c) & (d). In fact, Fig. 8(c) shows the $OC_{avg}$ depends linearly on $\langle |u_{\vec{k}_{RA}}|^2 \rangle / V_{SLR}$. On the other hand, in Fig. 8(d), the $OC_{avg}$ first increases rapidly with the Q-factor but begins to saturate at higher Q-factor. This is reasonable as the Q-factor is bound by $\frac{\Omega^2}{(\omega_{RA} - \tilde{\omega}_2)^2 + \Omega^2}$, in which further decrease in $\omega_{RA} - \tilde{\omega}_2$, or increase in the Q-factor, does not boost the $OC_{avg}$ anymore and the Lorentzian profile will reach 1 eventually.

## VII. CONCLUSION

In summary, we have developed rational schemes for producing superchiral SLRs from 3D bipartite metallic nanorod arrays. We first study the coupling constants between the LSP arising from nanoparticles and the diffractive RAs for exciting TE- and TM-SLRs. It is found the orientation of the nanorod lying in the plane normal to the propagation direction of the RA plays a major role in governing the couplings. TE- and TM-SLRs can be excited independently by using horizontal and vertical nanorods. Then, two orthogonally oriented nanorods are used to form the bipartite arrays. We analytically formulate the OC of the bipartite system and find it depends on the relative displacement between two nanorods, the near-field strength, and the interplay between the coupling constants and the Q-factor of the SLR resonance. By properly engineering these parameters, we have achieved the averaged OC of 27 times stronger than that of CPL over the entire system. Our work demonstrates the possibility of making large area, instead of spatially localized at a small region, chiral surface waves for facilitating chiral light-matter interaction, which will find applications in sensing, display, and communication.

## VIII. ACKOWLEDGEMENTS

This research was supported by the Chinese University of Hong Kong through Area of Excellence (AoE/P-02/12) and Innovative Technology Funds (ITS/133/19 and UIM/397).

spectroscopy: principles and applications, Nanoscale **13**, 581 (2021).

18. S. Yoo and Q. Park, Metamaterials and chiral sensing: a review of fundamentals and applications, Nanophotonics **8**, 249 (2019)

19. W. Ma, L. Xu, L. Wang, C. Xu, and H. Kuang, Chirality-based biosensors, Adv. Mater. **29**, 1805512 (2019)

20. J. A. Polo and A. Lakhtakia, Surface electromagnetic waves: A review, Laser & Photon. Rev. **5**, 234 (2011).

21. A. V. Zayats, I. I. Smolyaninov, and A. A. Maradudin, Nano-optics of surface plasmon polaritons, Phys. Rep. **408**, 131 (2005).

22. K. Y. Bliokh and F. Nori, Transverse and longitudinal angular momenta of light, Phys. Rep. **592**, 1 (2015).

23. A. Aiello, P. Banzer, M. Neugebauer, and G. Leuchs, From transverse angular momentum to photonic wheels, Nat. Photon. **9**, 789 (2015).

24. S. Yoo and Q. Park, Chiral light-matter interaction in optical resonators, Phys. Rev. Lett. **114**, 203003 (2015).

25. B. M. Maoz, Y. Chaikin, A. B. Tesler, O. Elli, Z. Fan, A. O. Govorov, and G. Markovich, Amplification of chiroptical activity of chiral biomolecules by surface plasmons, Nano Lett. **13**, 1203 (2013).

26. M. Hentschel, M. Schäferling, X. Duan, H. Giessen, N. Liu, Chiral plasmonics, Sci. Adv. **3**, e1602735 (2017).

27. Z. Fan and A. O. Govorov, Chiral nanocrystals: Plasmonic spectra and circular dichroism, Nano Lett. **12**, 3283 (2012).

28. J. T. Collins, C. Kuppe, D. C. Hooper, C. Sibilia, M. Centini, and V. K. Valev, Chirality and chiroptical effects in metal nanostructures: Fundamentals and current trends, Adv. Opt. Mater. **5**, 1700182 (2017).

29. X. Yin, M. Schäferling, B. Metzger, and H. Giessen, Interpreting chiral nanophotonic spectra: The plasmonic Born−Kuhn model, Nano Lett. **13**, 6238 (2013).

30. E. Hendry, T. Carpy, J. Johnston, M. Popland, R. V. Mikhaylovskiy, A. J. Lapthorn, S. M. Kelly, L. D. Barron, N. Gadegaard, and M. Kadodwala, Ultrasensitive detection and characterization of biomolecules using superchiral fields, Nat. Nano. **7**, 783 (2010).

31. Y. Chen, C. Zhao, Y. Zhang, and C. W. Qiu, Integrated molar chiral sensing based on high-Q
14

**FIGURE**

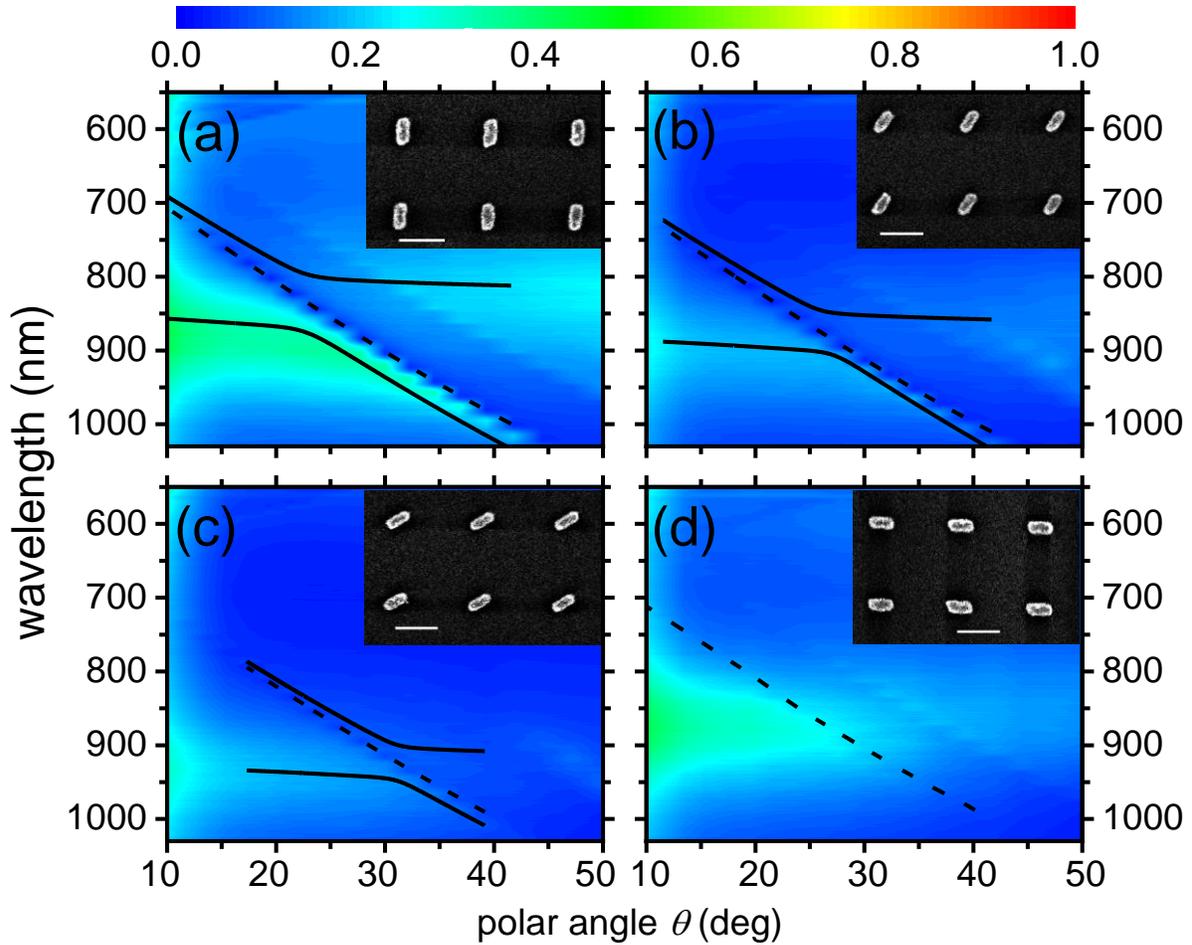

Figure 1. The $\theta$-resolved reflectivity mappings measured along the $\Gamma$-X direction for $\rho =$ (a) 0°, (b) 30°, (c) 60° and (d) 90° excited with a 45° linearly polarized light. The dashed lines are the ($-$1,0) RAs and the solid lines are the upper and lower coupled bands. The coupled bands are not present in (d). The insets show the SEM image of each sample of corresponding $\rho$, and the scale bar is 200 nm.

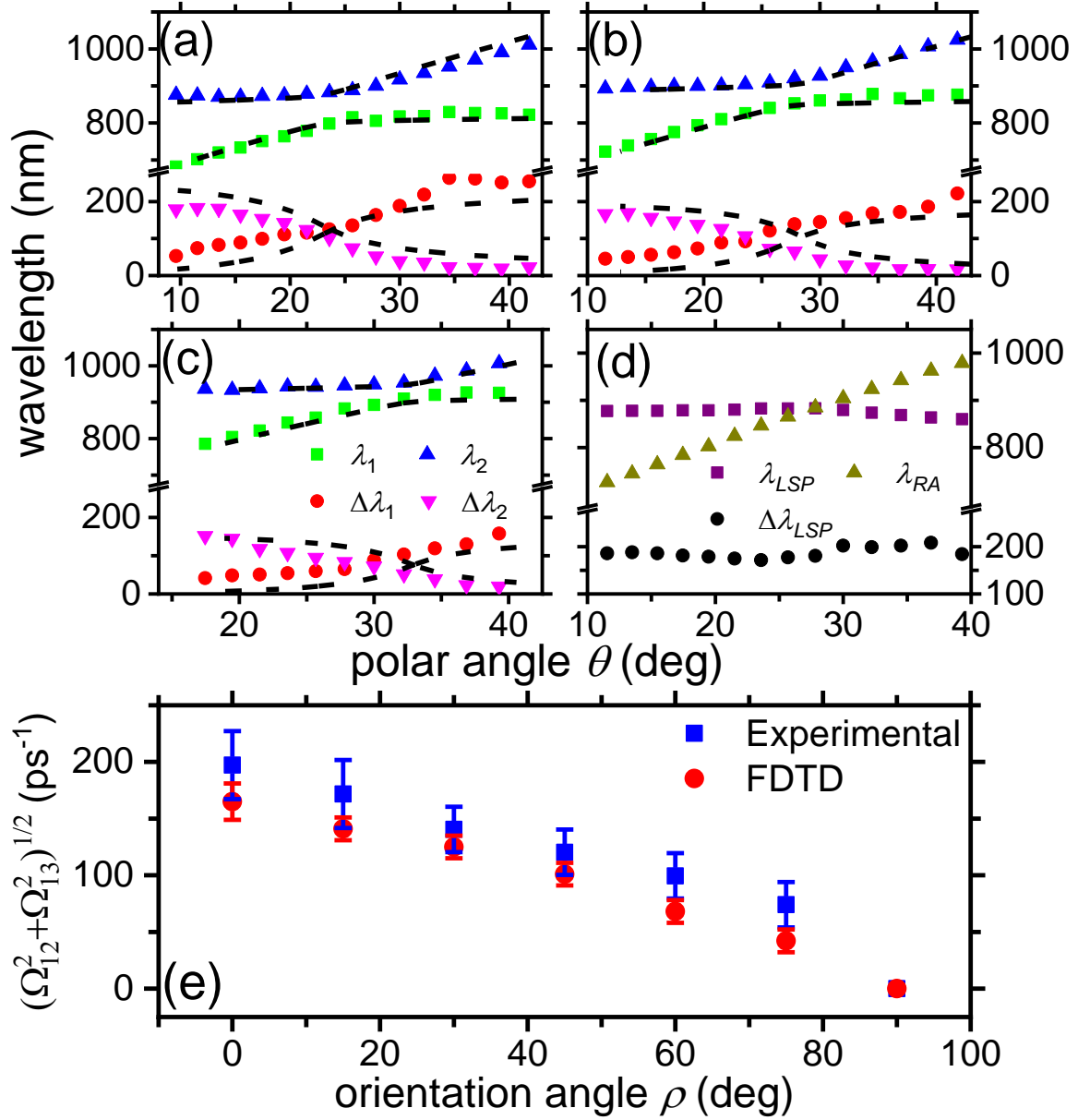

Figure 2. The plots of the spectral positions $\lambda_{1,2}$ and the linewidths $\Delta\lambda_{1,2}$ of the two bands for $\rho =$ (a) 0°, (b) 30°, and (c) 60° as a function of $\theta$. The dash lines are the best fits determined by temporal coupled mode theory. (d) The plots of the spectral positions $\lambda_{LSP,RA}$ and the linewidths $\Delta\lambda_{LSP}$ of the LSP and RA for $\rho = 90°$ array as a function of $\theta$. (e) The plot of experimental and FDTD simulated $\sqrt{\Omega_{12}^2 + \Omega_{13}^2}$ as a function of $\rho$.

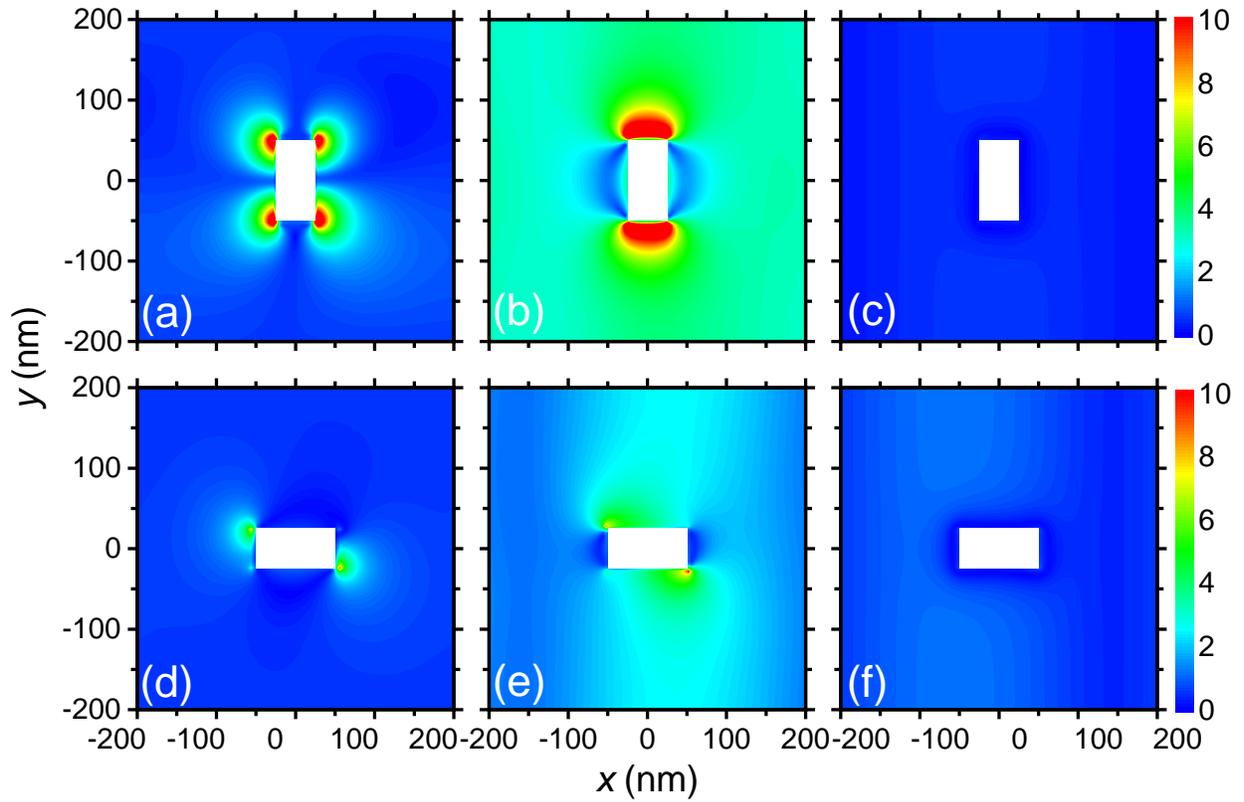

Figure 3. The near-field patterns of (–1,0) SLR with (a) – (c) $\rho = 0°$ and (d) – (f) $\rho = 90°$ using 45° linearly polarized incident source at $\lambda = 950$ nm. The *x*-components are illustrated in (a) and (d), *y*-components in (b) and (e), and *z*-components in (c) and (f). The blank areas are the nanorods.

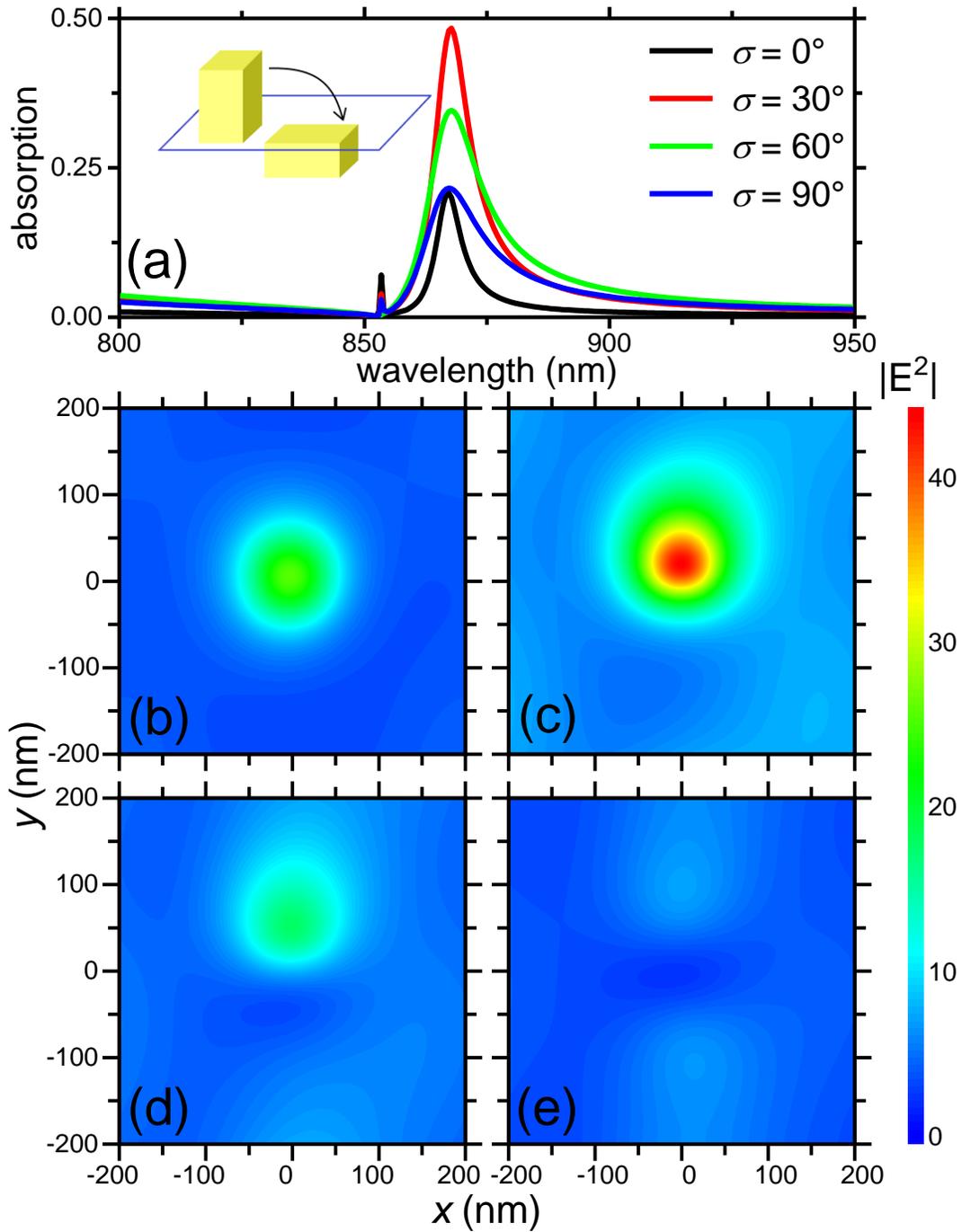

Figure 4. (a) The absorption spectra of the arrays for $\sigma = 0°$, $30°$, $60°$, and $90°$. The inset shows the nanorod is tilted from $\sigma = 0°$ to $90°$. The SLR near-field intensity patterns for $\sigma =$ (b) $0°$, (c) $30°$, (d) $60°$, and (e) $90°$, at 100 nm above the surface at $\lambda = 868$ nm.

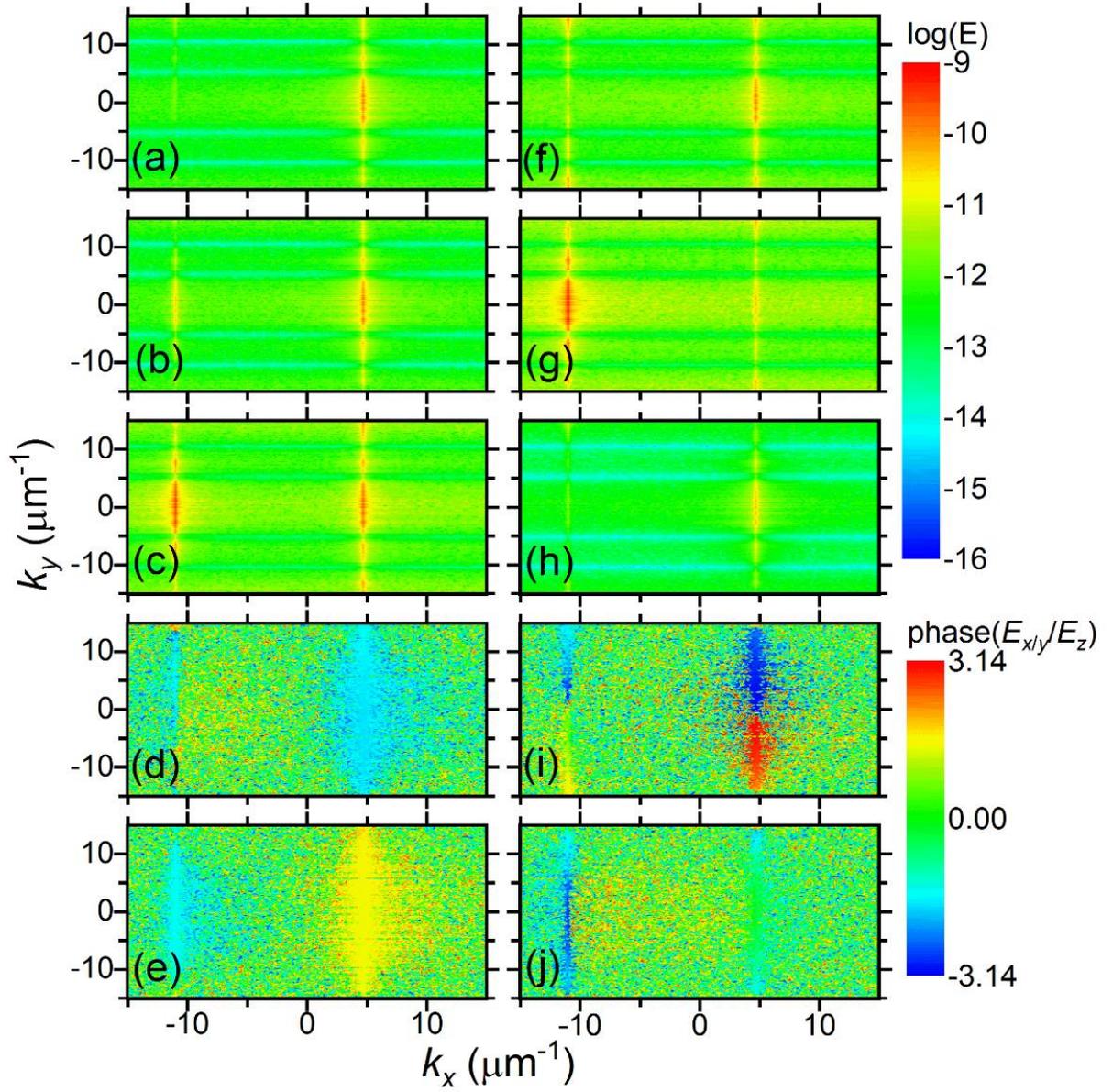

Figure 5. The magnitude (in log scale) and the phase difference mappings of SLR near-field pattern for $\sigma =$ (a) – (e) 0° and (f) – (j) 90° arrays in momentum space. The *x*-components are illustrated in (a) and (f), *y*-components in (b) and (g), and *z*-components in (c) and (h). The phase difference between *x*- and *z*-component are illustrated in (d) and (i), and between *y*- and *z*-component in (e) and (j).

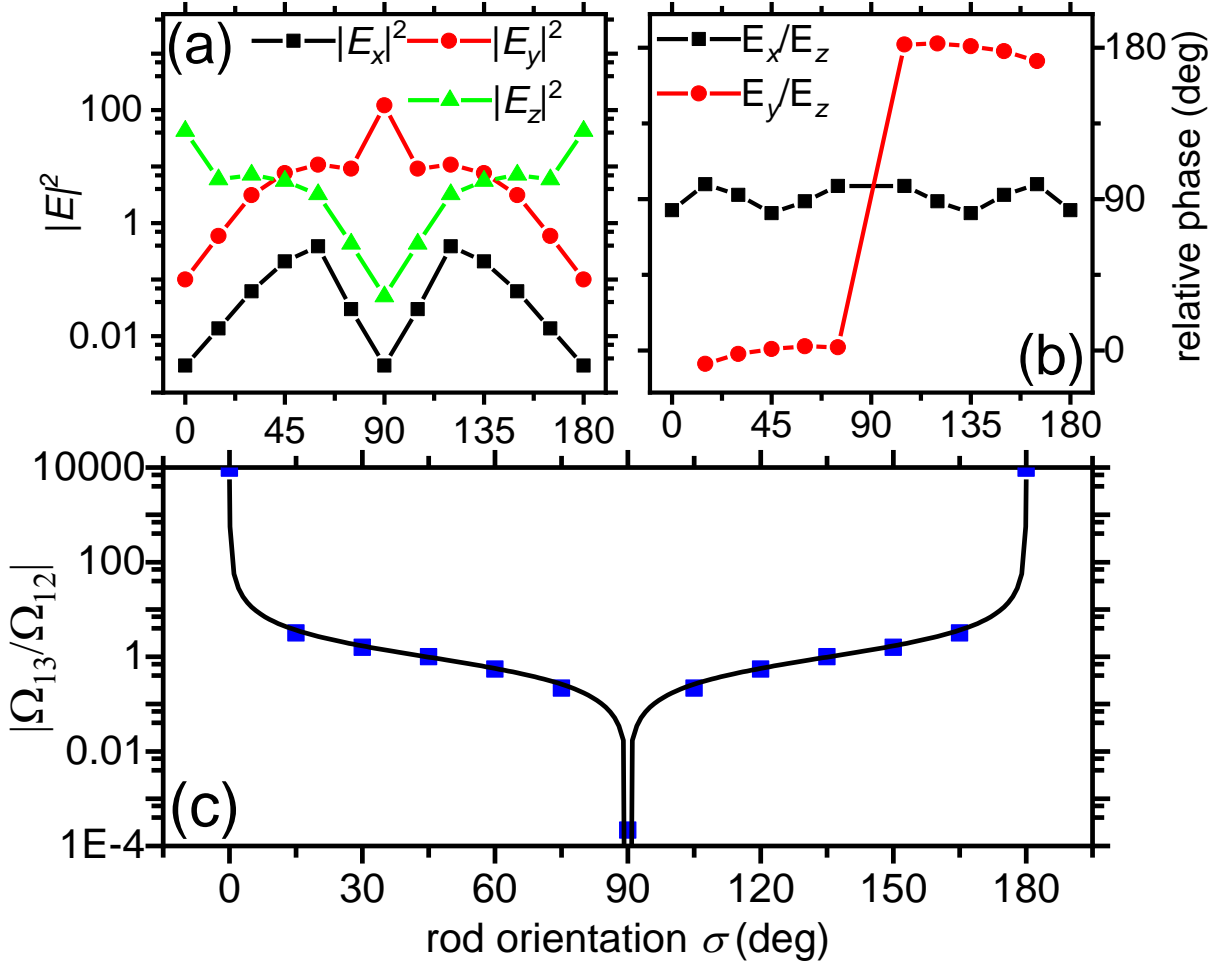

Figure 6. (a) The plot of the Bloch $|\vec{E}_x|^2$, $|\vec{E}_y|^2$ and $|\vec{E}_z|^2$ (in log scale) as a function of σ. (b) The plot of the phase differences between the field components as a function of σ. (c) The plot of $|\Omega_{TM}/\Omega_{TE}|$ ratio (in log scale) as a function of σ. The solid line is the best fit determined by $|\beta \cot \sigma|$.

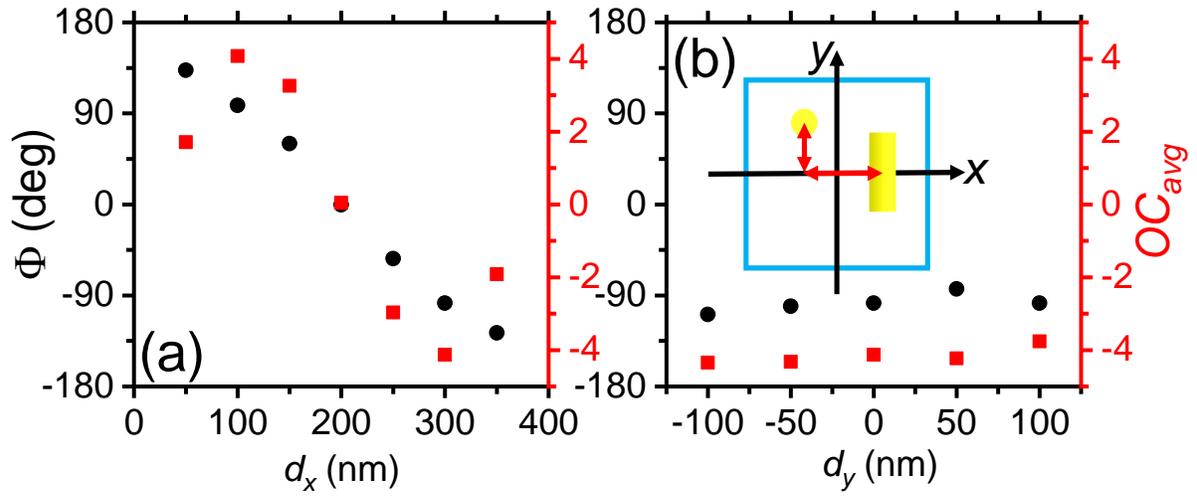

Figure 7. The plots of the phase difference Φ between $E_y$ and $E_z$ (black circles) and the normalized $OC_{avg}$ (red squares) as a function of (a) $d_x$ and (b) $d_y$.

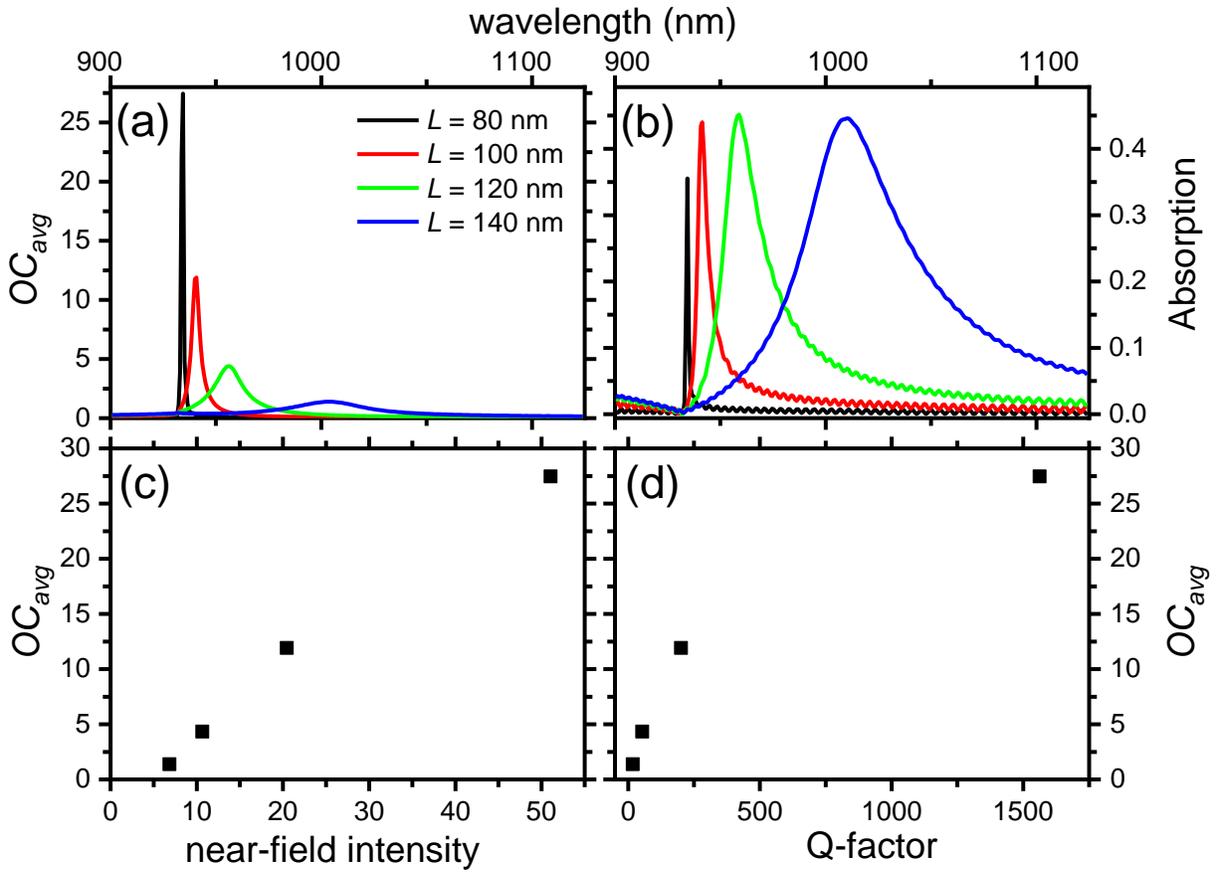

Figure 8. (a) The $OC_{avg}$ spectra and (b) the absorption spectra of (–1,0) SLRs with $L = 80 – 140$ nm. The plots of $OC_{avg}$ ratio against (c) averaged near-field intensity and (d) the Q-factor of SLR.

# Supplementary Information

# Realization of superchiral surface lattice resonances in three-dimensional bipartite nanoparticle arrays

Joshua T.Y. Tse and H.C. Ong

Department of Physics, The Chinese University of Hong Kong, Shatin, Hong Kong, People's Republic of China

**A. Scanning electron microscopy images of monopartite nanorod arrays**

Figure S1 shows the plane-view scanning electron microscopy (SEM) images of 2D Au rectangular cuboid nanorod arrays azimuthally rotated, with respect to the Γ-Y direction, from $\rho = 0°$ to $90°$ with a step size = 15°.

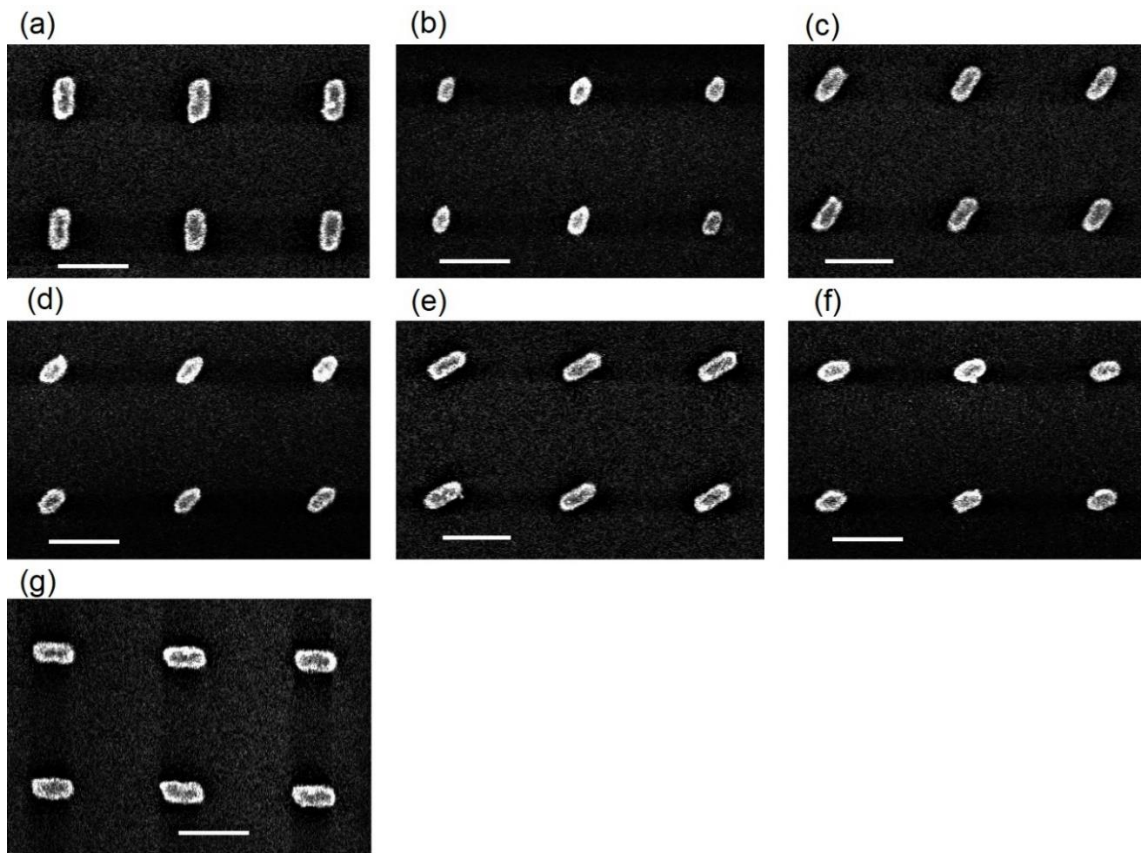

Fig. S1. The SEM images of the nanorods with $\rho$ = (a) 0°, (b) 15°, (c) 30°, (d) 45°, (e) 60°, (f) 75°, and (g) 90°.

## B. Experimental angle-resolved reflectivity mapping of monopartite nanorod arrays

The 45° linearly polarized angle-resolved reflectivity mappings of 2D Au rectangular cuboid nanorod arrays azimuthally rotated from $\rho$ = 0° to 90° with a step size = 15° taken along the Γ-X direction. The dash lines are the RA calculated by using the phase matching equation $(n/\lambda)^2 = (n\sin\theta/\lambda - 1/P)^2$, where $n$ = 1.5 is the refractive index of the environment. We observe an avoided crossing at $\theta \approx 23.6° - 30°$ due to the coupling between the LSP and the RA, forming an energy band gap as well as two upper and lower bands, as shown by the solid lines. For $\rho$ = 90°, the gap closes, indicating no coupling between LSP and the RA.

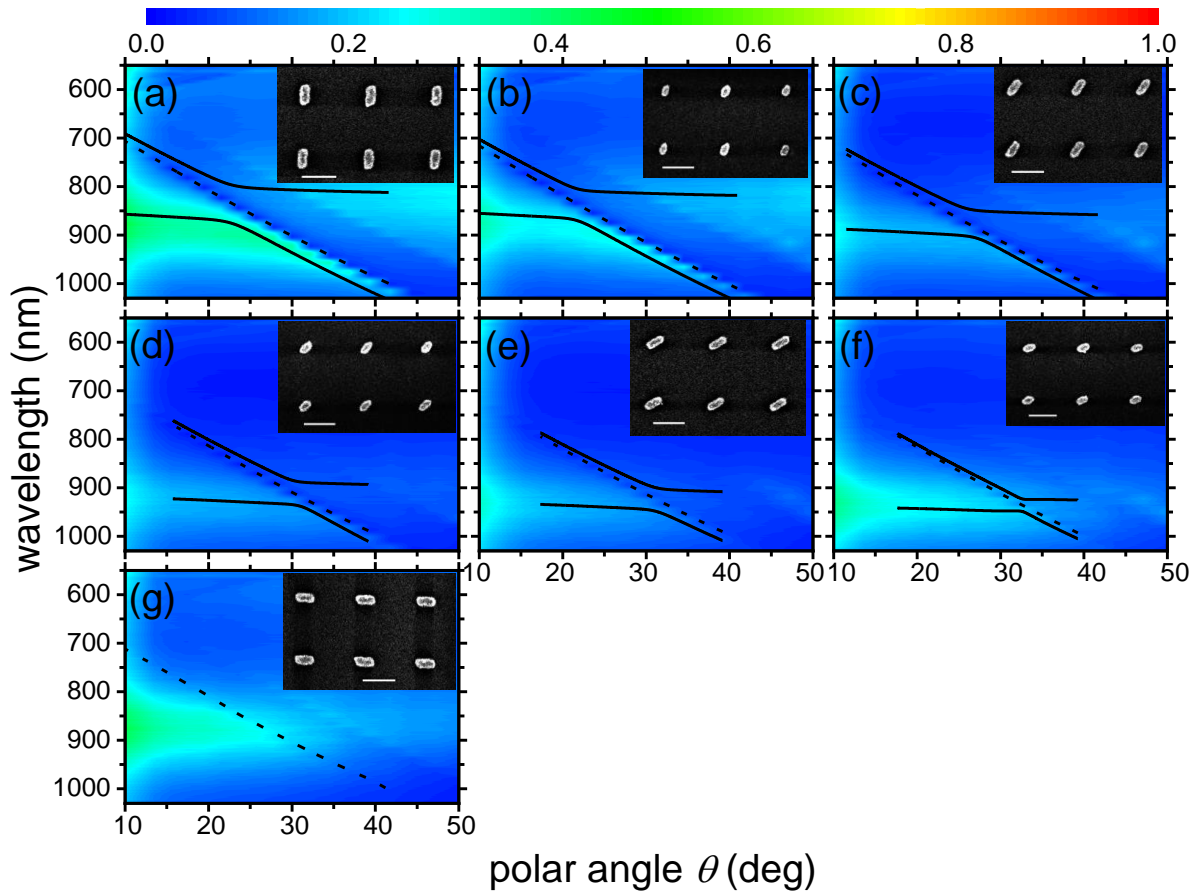

Fig. S2. The angle-resolved reflectivity mappings of the nanorods with $\rho$ = (a) 0°, (b) 15°, (c) 30°, (d) 45°, (e) 60°, (f) 75°, and (g) 90°. The corresponding SEM images are shown in the insets.

## C. Experimental $\lambda_{1,2}$ and $\Delta\lambda_{1,2}$ of monopartite nanorod arrays

The spectral positions $\lambda_{1,2}$ and the linewidths $\Delta\lambda_{1,2}$ of the upper and lower bands for $\rho$ from 0° to 90° with a step size of 15° are plotted in Fig. S3 as a function of polar angle $\theta$. The dashed lines are the best fit by CMT for determining $\sqrt{\Omega_{12}^2 + \Omega_{13}^2}$.

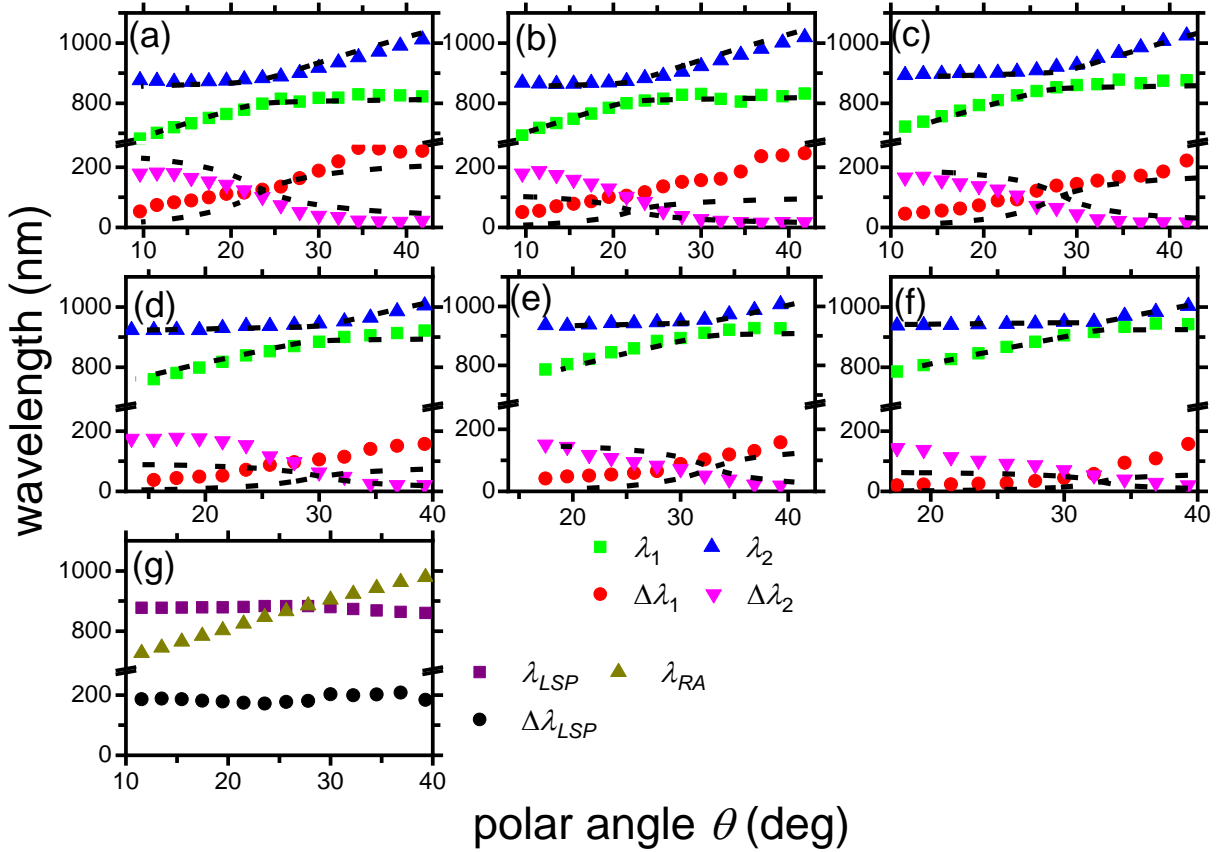

Fig. S3. The plots of the spectral positions $\lambda_{1,2}$ and the linewidths $\Delta\lambda_{1,2}$ of the upper and lower bands for $\rho$ = (a) 0°, (b) 15°, (c) 30°, (d) 45°, (e) 60°, and (f) 75° arrays as a function of $\theta$. The black dash lines are the best fits for determining $\sqrt{\Omega_{12}^2 + \Omega_{13}^2}$. (g) The plots of the spectral positions $\lambda_{LSP,RA}$ and linewidths $\Delta\lambda_{LSP}$ of the LSP and RA for $\rho = 90°$ as a function of $\theta$.

### D. FDTD simulations on monopartite nanorod arrays

We have performed FDTD simulations on 2D Au monopartite nanorod arrays to verify the experimental results. Fig. S4(a) shows the unit cell for the simulations. The cuboid nanorod has length = 100 nm, width = 50 nm, and height = 50 nm. The square lattice period is 400 nm. Fig. S4(b) – (e) show the 45° linearly polarized angle-resolved reflectivity mappings with different azimuthal rotation angles $\rho$ taken along the Γ-X direction. The dispersive (–1,0) RAs are overlayed as the dashed line and the upper and lower bands are shown by the solid lines. Fig. 4(f) – (h) show the spectral position $\lambda_{1,2}$ and the linewidth $\Delta\lambda_{1,2}$ plots together with the best fits determined by using CMT, for the arrays $\rho = 0°$, $30°$ and $60°$. Fig. S4(i) shows the spectral position $\lambda_{LSP,RA}$ and the linewidth $\Delta\lambda_{LSP}$ plots at $\rho = 90°$, indicating the LSP and RA are not coupling. The simulation results agree with the experiment.

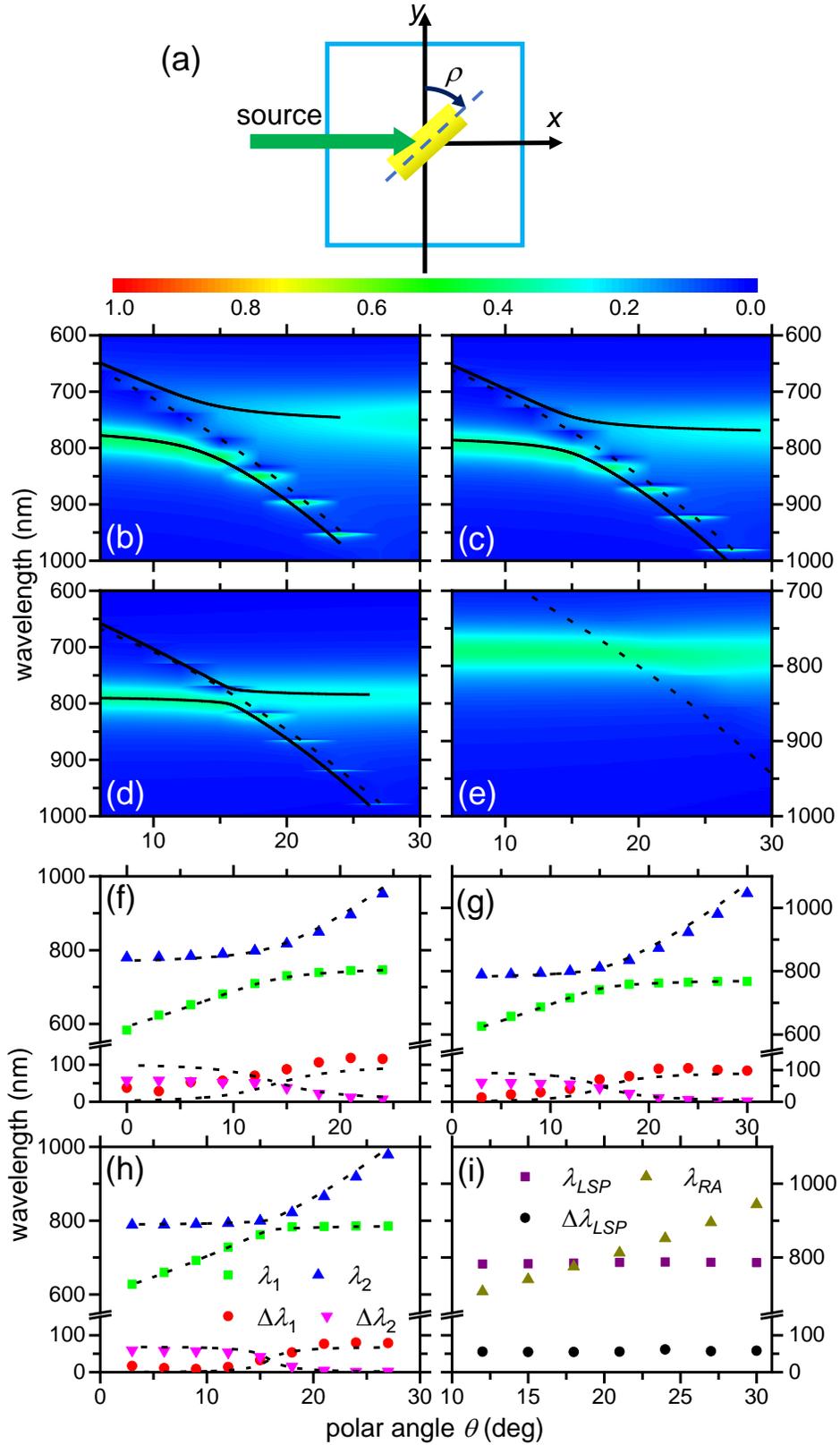

Fig. S4. (a) The FDTD unit cell. (b) – (e) The FDTD angle-resolved reflectivity mappings for $\rho$ = (b) 0°, (c) 30°, (d) 60°, and (e) 90°. The dispersive (–1,0) RAs are overlayed as the dashed line

and the upper and lower bands are shown by the solid lines. (f) – (h) The $\lambda_{1,2}$ and $\Delta\lambda_{1,2}$ plots (symbols) together with the best fits (dash lines) determined by using CMT for $\rho =$ (f) 0°, (g) 30° and (h) 60°. (i) The plot of the $\lambda_{LSP,RA}$ and $\Delta\lambda_{LSP}$ at $\rho = 90°$.

### E. FDTD simulated near-field patterns of monopartite nanorod arrays

We have simulated the field patterns of the (–1,0) SLRs for $\rho = 0°$, 30°, 60°, and 90° arrays taken at $\theta = 24°$ along the Γ-X direction. While the field patterns for $\rho = 0°$ and 90° have been provided in the main text, the results for $\rho = 30°$ and 60° are shown in Fig. S5. We clearly see the SLRs are TE-like for both $\rho$.

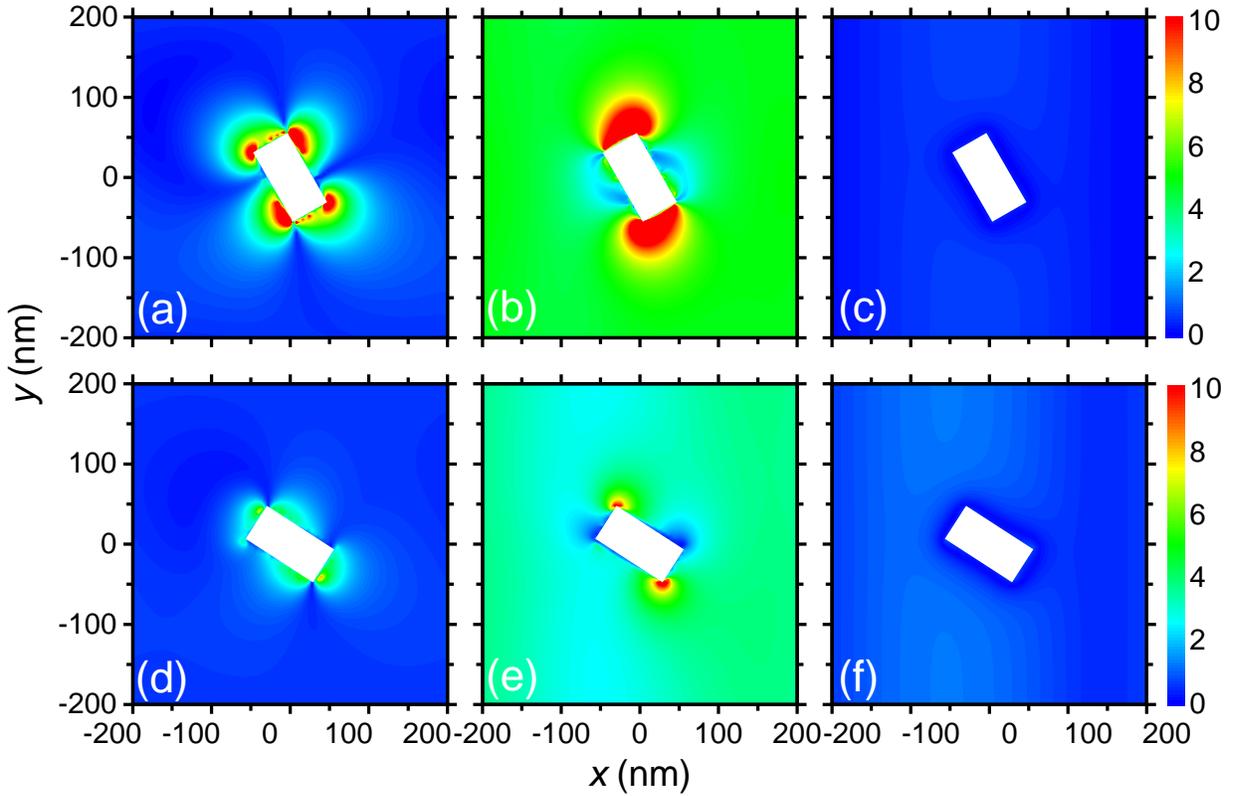

Fig. S5. The near-field patterns of SLR with (a) – (c) $\rho = 30°$ and (d) – (f) $\rho = 60°$ using 45° linearly polarized incident source. The x-components are illustrated in (a) and (d), y-components in (b) and (e), z-components in (c) and (f). The blank areas are the nanorods.

### F. Fourier transformed results for $\sigma = 30°$ and 60° arrays

While the Fourier transformed results for $\sigma = 0°$ and 90° arrays are provided in the main text, the results for $\sigma = 30°$ and 60° arrays are shown here. Fig. S6(a) – (e) and (f) – (j) show the magnitude and relative phase mappings in momentum space for $\sigma = 30°$ and 60° arrays.

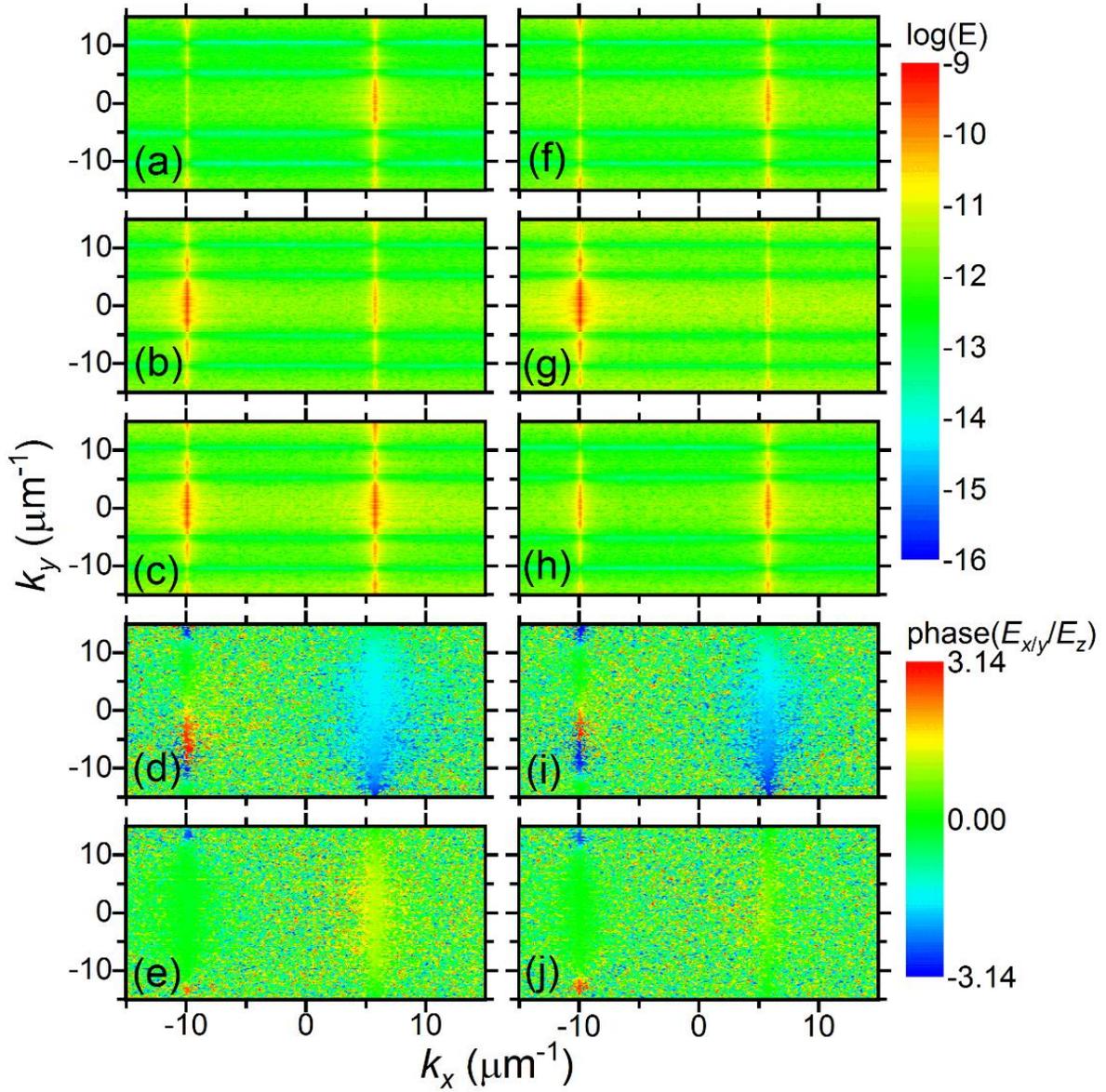

Fig. S6. The magnitude (in log scale) and relative phase mappings of SLR near-field pattern for $\sigma =$ (a) – (e) 30° and (f) – (j) 60° in momentum space. The $x$-components are illustrated in (a) and (f), $y$-components in (b) and (g), and $z$-components in (c) and (h). The phase difference between $x$- and $z$-component are illustrated in (d) and (i), and between $y$- and $z$-component in (e) and (j).

## G. Derivation of the averaged OC

While the total electric field is given in Eq. (8), the total magnetic field is:

$$\vec{B} = \frac{n}{c}\sqrt{\frac{2}{n^2\varepsilon_o}}\left(\frac{\beta^{TE}}{\sqrt{V_2^{eff,TE}}}\Omega_{TE} u^{TE}_{\vec{k}_{RA}} e^{i\vec{k}_{RA}\cdot\vec{r}}\left(i\sinh\tau\hat{x}+\cosh\tau\hat{z}\right) - \frac{\beta^{TM}}{\sqrt{V_2^{eff,TM}}}\Omega_{TM} u^{TM}_{\vec{k}_{RA}} e^{i\vec{k}_{RA}\cdot\vec{r}}\hat{y}e^{i\Phi}\right), \quad \text{(S1)}$$

given by $\vec{B} = \frac{1}{\omega}\vec{k}\times\vec{E} = \frac{n}{c}\hat{k}\times\vec{E}$ where $\vec{k} = \vec{k}_{RA} - i\hat{z}\sqrt{\left(\frac{2\pi}{\lambda_{RA}}\right)^2 - \left(\frac{2\pi}{\lambda}\right)^2}$. The OC can then be calculated by using Eq. 2 to give:

$$OC = \frac{2\omega}{nc}\text{Im}\left[\left(\frac{u^{TE}_{\vec{k}_{RA}}}{\sqrt{V_2^{eff,TE}}}\beta^{TE}\Omega_{TE}\right)^*\left(\frac{u^{TM}_{\vec{k}_{RA}}}{\sqrt{V_2^{eff,TM}}}\beta^{TM}\Omega_{TM}\right)e^{i\Phi}\right]. \tag{S2}$$

Since $\Omega_{TE} \propto \hat{p}\cdot\hat{y}$ and $\Omega_{TM} \propto \hat{p}\cdot\hat{z}$, the identical nanorods give $\Omega_{TE} = \Omega_{TM} = \Omega$ and $\beta^{TE}\Omega_{TE} = \beta^{TM}\Omega_{TM} = \frac{\Omega}{\sqrt{(\omega_{RA}-\tilde{\omega}_2)^2+\Omega^2}}$. Assuming the spatial profile of TE- and TM-SLR are similar, we approximate $V_2^{eff,TE} \approx V_2^{eff,TM} = V_{SLR}$ and $|u^{TM}_{\vec{k}_{RA}}| \approx |u^{TE}_{\vec{k}_{RA}}| = |u_{\vec{k}_{RA}}|$. Therefore, we derive the averaged OC to be:

$$OC_{avg} = \frac{\omega}{nc}\frac{\langle|u_{\vec{k}_{RA}}|^2\rangle}{V_{SLR}}\frac{\Omega^2}{(\omega_{RA}-\tilde{\omega}_2)^2+\Omega^2}\sin\Phi. \tag{S3}$$

## H. FDTD simulated OC pattern for $L$ = 80 nm 3D bipartite nanorod array

At $\lambda$ = 934 nm, we have simulated the OC ratio mapping for $L$ = 80 nm bipartite nanorod array in Fig. S7. The strongest OC ratio is 102.5 whereas the averaged OC ratio is 27.5.

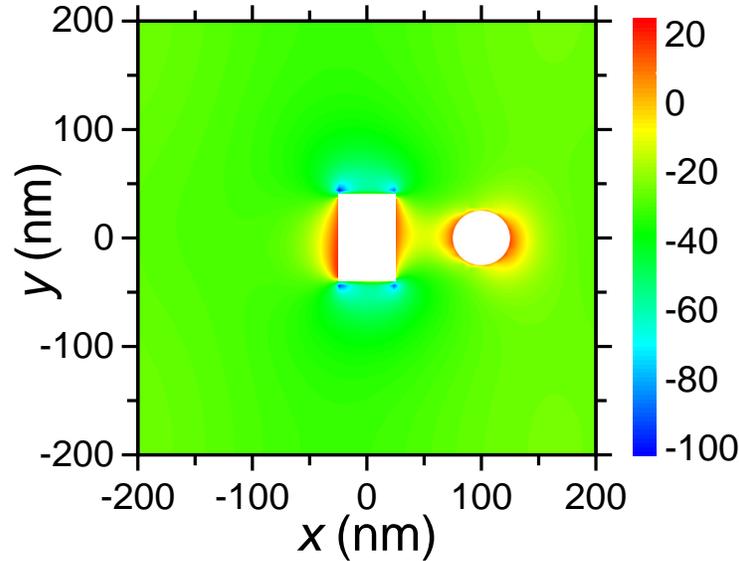

Fig. S7. The OC ratio mapping for $L$ = 80 nm bipartite nanorod array. The blank areas are occupied by the nanorods.